\documentclass[usenatbib]{aa}
\usepackage[varg]{txfonts}

\usepackage{graphicx}
\usepackage{txfonts}
\usepackage{amsmath}	
\usepackage{amssymb,multirow}	

\usepackage[colorlinks=true,
linkcolor=blue,
citecolor=blue,
urlcolor=blue,
plainpages=true        
]{hyperref}     %

\newcommand{\dd}[2]{\frac{\mathrm{d} #1}{\mathrm{d} #2}}

\newcommand{\mi}{\mathrm{i}}
\newcommand{\me}{\mathrm{e}}

\newcommand{\rin}{r_{\mathrm{in}}}
\newcommand{\rout}{r_{\mathrm{out}}}

\newcommand{\Msun}{M_{\odot}}

\newcommand{\Mj}{M_{\rm J}}

\begin{document} 

\titlerunning{TTV Signature of Planet Migration in K2-24}

\title{Transit-Timing Variation Signature of Planet Migration: The Case of K2-24}

\subtitle{}

\author{Jean Teyssandier\thanks{E-mail: jean.teyssandier@unamur.be} and Anne-Sophie Libert
}

\institute{naXys, Department of Mathematics, University of Namur, Rempart de la Vierge 8,
5000 Namur, Belgium}

\date{Accepted XXX. Received YYY; in original form ZZZ}

\abstract{The convergent migration of two planets in a gaseous disc can lead to capture in mean motion resonance (MMR). In addition, pairs of planets in or near MMRs are known to produce strong transit timing variations (TTVs). In this paper we study the impact of disc-induced migrations on the TTV signal of pairs of planets that enter a resonant configuration. We show that disc-induced migration creates a correlation between the amplitude and the period of the TTVs. We study the case of K2-24, a system of two planets whose period ratio indicates that they are in or near the 2:1 MMR, with non-zero eccentricities and large-amplitude TTVs. We show that a simple disc-induced migration cannot reproduce the observed TTVs, and we propose a formation scenario in which the capture in resonance occurring during migration in a disc with strong eccentricity damping is followed by eccentricity excitation during the dispersal of the disc, assisted by a third planet whose presence has been suggested by radial velocity observations. This scenario accounts for the eccentricities of the two planets and their period ratio, and accurately reproduces the amplitude and period of the TTVs. It allows for a unified view of the formation and evolution history of K2-24, from disc-induced migration to its currently observed properties.}



\maketitle

\section{Introduction}
\label{sec:intro}

Pairs of adjacent planets in the\textit{ Kepler} catalogue which are close to a MMR are often found to be exterior to the exact commensurability\footnote{In this paper, we define the ``exact commensurability'' as the ratio of two integers. Hence the exact commensurability associated with the 2:1 MMR is 2. Broadly speaking, planet pairs whose orbital period ratios fall near the exact commensurability are refered to as being ``near'' the resonance, regardless of whether or not they are resonant in the dynamical sense.}, with a dearth of planets interior to it \citep{fabrycky12}. The fact that pairs of planets can be found in or near MMR is not a surprise: convergent migration of two planets in a disc is known to be able to trap them in a MMR \citep[see, e.g.,][]{snellgrove01,lp02}, and the ratio of their orbital periods is expected to be very close to the exact commensurability. However, the fact that these pairs of planets are found preferably on one side of the exact commensurability is considered more surprising.

Several studies \citep[see, e.g.,][]{novak03,papaloizouterquem10,lw12,bm13,dl14} have argued that tides raised on the planets can cause a resonant repulsion of the two planets, hereby slightly increasing their period ratios and explaining the trend seen in the \textit{Kepler data}. Other mechanisms have been proposed that also favour planets pairs lying just outside of exact commensurabilities: In-situ growth of planets via planetesimal accretion \citep{petrovich13} or interactions with a disc of planetesimals \citep{chatterjee15}.

However it is worth noting that one should not expect resonant pairs to be observed exactly at the exact commensurability. Precession of the orbits tends to shift the centre of the resonance away from its exact commensurability. For convergent migration, this offset can naturally produce pairs of planets with period ratios larger than the exact commensurability \citep[see, e.g.,	][]{ramos17,tp19}.

In parallel to their period ratio distribution, systems near MMRs have also drawn a lot attention due to the significant TTVs they often produce. These variations result from perturbations (in general gravitational perturbations from other planets), which modify the  periodicity at which a planet transits in front of its parent star. In addition to being able to reveal or confirm the presence of non-transiting planets \citep{miralda02,schneider04,agol05,hm05,cochran11,ford12,steffen12a}, TTVs provide a way of inferring masses and eccentricities in multiple-transiting systems \citep{nesvorny08,lithwick12,hadden17}. In particular, systems near MMRs are subject to large TTV signals \citep[see, e.g.,][]{lithwick12}. The amplitude and period of the TTV signal depends strongly on the distance to the exact commensurability, and on the eccentricities of the planets. These quantities are often shaped during the phase of planet-disc interactions.

Since disc migration can produce planets in or near MMRs (a process which tends to excite their eccentricities), and since these quantities play a key role in TTVs of planet pairs, it is natural to ask what is the effect of disc migration on TTV signals. This is the goal of this work. We focus on the K2-24 system, for which \citet{petigura18} obtained accurate periods, masses and TTVs, and managed to estimate the eccentricities. In Section~\ref{sec:k224}, we review in more details why K2-24 is a good laboratory to test the impact of planet migration on TTVs. In Section~\ref{sec:disc}, we carry out numerical simulations to study the TTV signals for pairs of planets that have migrated in a disc. We then suggest that an additional mechanism involving a third planet and an evaporating disc can explain the current architecture of the K2-24 system (Section~\ref{sec:disc3p}). Finally we discuss the implications of our results in Section~\ref{sec:discussion} and conclude in Section \ref{sec:conclusion}.

\section{Generalities}
\label{sec:k224}

\subsection{Properties of K2-24}
\label{sec:review}

The planetary system K2-24 was first reported by \citet{petigura16}, with subsequent refined measurements by \citet{dai16}, \citet{sinukoff16}, \citet{crossfield16} and \citet{mayo18}. The latest data for this system are provided by \citet{petigura18}, who conducted a joint TTV and radial velocity (RV) analysis which allowed them to better constrain the masses and orbits of K2-24b and c. We summarize their main findings below.

The K2-24 system has at least two planets (noted b and c), with masses $m_b=19M_{\oplus}$, $m_c=15M_{\oplus}$, and periods $20.89$ and $42.34$  days (period ratio of 2.028), around a star of mass $M_*=1.07\Msun$. The system is thus near the 2:1 MMR and shows strong TTVs. The eccentricities of the two known planets are non-zeros. In particular, RV data only ruled out high eccentricity orbits, giving upper limits of $e_{\rm b} < 0.39$ and $e_{\rm c} < 0.34$. On the other hand, TTV data gave a constraint on a linear combination of masses and eccentricities. Using RV data to break the mass degeneracy, \citet{petigura18} managed to get a range of $e_{\rm b}$ and $e_{\rm c}$  consistent with their joint RV/TTV analysis (see their Figure 7).
In an attempt to better constrain the eccentricities, \citet{petigura18} assumed that K2-24 is part of the population of TTV-active Kepler multi-planet systems, and used a Rayleigh eccentricity distribution with $<e> = 0.03$ as a prior \citep[see, e.g.,][]{wu13}. From that they deduced the eccentricities of K2-24b and c to be around 0.07.
It is worth noting that relaxing this assumption allows for best fits with eccentricities in the range 0.1--0.2 \citep[see Figure 7 of][]{petigura18}. The two key features of the TTV signal are its amplitude of about 0.25 and 0.5 days for planet b and c, respectively, and its period of $\sim$ 1580 days.

In addition, \citet{petigura18} reported the possible detection of K2-24d, a third planet with mass $54^{+14}_{-14}~M_{\oplus}$ and semi-major axis $1.15^{+0.06}_{-0.05}~\text{au}$.

In order to understand why the specific characteristics of K2-24 are at odds with a simple model of disc migration, it is useful to first review the theory behind TTV signals near resonances, which we do in the next section.

\subsection{Theory of TTVs}
\label{sec:ttv}
TTVs have been recognized for a long time as a powerful technique to detect unseen planets in systems with at least one transiting planet, and to better constrain masses and eccentricities \citep{miralda02,agol05,hm05,agol18}.

As previously mentioned, K2-24b and c exhibit large amplitude TTVs, with a period of $\sim 1580$ days. In order to use these TTV measurements as a constraint on the system's formation and dynamical history, it is useful to review the basis of the theory of TTVs. In particular, \citet{lithwick12} developed an analytical model for two planets near a first-order MMR, as is the case for K2-24, which is near the 2:1 resonance. We summarize the main results of this analysis here.

First, let us focus on the period of the TTV signal. We consider two planets near the 2:1 MMR with orbital periods $P_1$ and $P_2$ such as $P_1<P_2$, and masses $m_1$ and $m_2$. Let us define the normalized distance to the exact commensurability as
\begin{equation}
\label{eq:delta}
\Delta\equiv\frac{1}{2}\frac{P_2}{P_1}-1.
\end{equation}
Close to a MMR, \citet{steffen06} and \citet{lithwick12} showed that the TTV signal undergoes periodic cycles on a period given by the ``super-period'' $\mathcal{P}$, whose value depends on the distance to the exact commensurability:
\begin{equation}
\label{eq:ttv_period}
\mathcal{P}=\frac{P_2}{2\Delta}.
\end{equation}
For K2-24, we have $\Delta=0.0134$ and $\mathcal{P}=1580~\text{days}$. 

Now we turn to consider the amplitude of the TTV. \citet{lithwick12} showed that the key quantity to consider is the free eccentricity of the planets. For a planet with eccentricity $e$ and argument of pericentre $\varpi$, it is convenient to introduce the complex eccentricity $z=e \exp{(\mi \varpi)}$. The complex eccentricity can be decomposed into two terms, a \textit{free} eccentricity $z_{\rm free}$, and a \textit{forced} eccentricity $z_{\rm forced}$ \citep[see, e.g.,][]{md99}:
\begin{equation}
z=z_{\rm free}+z_{\rm forced}.
\end{equation}
The forced eccentricity arises from the resonant perturbations and is determined by the planets' proximity to exact commensurability, while the free eccentricity is unrelated to the resonance. Near the 2:1 resonance, \citet{lithwick12} showed that, to leading order in eccentricity, the forced eccentricity can be expressed as:
\begin{align}
\begin{pmatrix}
z_{\rm 1, forced} \\
z_{\rm 2, forced}
\end{pmatrix}
=
-\frac{1}{2\Delta}
\begin{pmatrix}
\mu_2 ~f ~(P_2/P_1)^{1/3} \\
\mu_1 ~ g 
\end{pmatrix}
e^{\mi \lambda}.
\end{align}
Here $\mu_i=m_i/M_*$, and $\lambda=2\lambda_2-\lambda_1$ is the longitude of conjunctions ($\lambda_1$ and $\lambda_2$ being the mean longitudes). In addition, $f$ and $g$ are coefficients that represent the secular and resonant coupling of the two planets. They can be found in Table~3 of \citet{lithwick12}. Near the 2:1 resonance, they read $f=-1.19+2.20\Delta$ and $g=0.4284-3.69\Delta$, at first order in $\Delta$. The main result of the analysis conducted by \citet{lithwick12} is that the key quantity in determining the amplitude of the TTV signal is a linear combination of the free complex eccentricity of each planet:
\begin{equation}
\label{eq:zfree}
Z_{\rm free}\equiv f~ z_{1,\rm free} + g~ z_{2,\rm free}.
\end{equation}
\citet{lithwick12} showed that the amplitude of the TTVs of each planet increases linearly with $|Z_{\rm free}/\Delta|$.

The TTV analysis of \citet{petigura18}, in synergy with their RV measurements, gave the following constraints for $Z_{\rm free}$ in K2-24:
\begin{align}
\label{eq:zfreek224}
\begin{split}
\text{Re}(Z_{\rm free})&=0.038^{+0.004}_{-0.003},\\
\text{Im}(Z_{\rm free})&=0.070^{+0.008}_{-0.007}.
\end{split}
\end{align}
As previously said, \citet{petigura18} then assumed a prior on the eccentricity distribution (a Rayleigh distribution parametrized by a mean eccentricity $\langle e \rangle = 0.03$), and deduced that $e_1\sim 0.06$ and $e_2\lesssim 0.07$. However, if one relaxes the assumption that the eccentricities in K2-24 follow such Rayleigh distribution, the only strong constraint on the eccentricities is given by Eq.~ (\ref{eq:zfreek224}). The contour of eccentricities allowed by the joint RV-TTV analysis can be seen in Figure 7 of \citet{petigura18}. Apart from the fact that none of the eccentricities are zero, we are left with a wide range of possible eccentricities for K2-24b and c.

\subsection{Departure from exact commensurability}
\label{sec:res}
Now that we are equipped with a good understanding of the TTV signal near the 2:1 resonance, we need to focus on the values of $\Delta$ (i.e., how far from the exact commensurability) which are expected from convergent disc migration.

The two resonant angles of a system in a 2:1 MMR are \citep[see, e.g.,][]{md99}
\begin{align}
\label{eq:res_angles}
\theta_{1,2}=2\lambda_2-\lambda_1-\varpi_{1,2}.
\end{align}
Let us consider a resonant state in which $\theta_{1}$ and $\theta_{2}$ librate and their time derivatives average to zero. In addition, without exterior perturbation, the two planets are locked in common apsidal precession, so that $\dot{\varpi}_1=\dot{\varpi}_2$. Setting $\dot{\theta}_{1,2}$ to zero, we arrive at
\begin{equation}
\label{eq:resloc}
\dot{\varpi}_1 \texttt{\,=\,} 2n_2-n_1,
\end{equation}
where $n_1$ and $n_2$ are the mean motions. Near the 2:1 resonance, at lowest order in eccentricity, $\varpi_1$ varies as \citep{md99}:
\begin{align}
\dot{\varpi}_1=n_2\frac{m_2}{M_*} \alpha^{-1/2} f_1\frac{1}{e_1}\cos\theta_1,
\end{align}
where $f_1$ can be expressed as Laplace coefficients, and takes the value $f_1\simeq-1.19$ for $\alpha=a_1/a_2=2^{-2/3}$. Finally \citep[in the most common case where $\theta_1$ librates around 0; see, e.g.,][]{lee04}, Eq.~(\ref{eq:resloc}) can be re-written:
\begin{equation}
\label{eq:resdist}
\Delta \texttt{\,=\,} 0.75\frac{m_2}{M_*}\frac{1}{e_1},
\end{equation}
where $\Delta$ is defined in Eq.~(\ref{eq:delta}). A similar result was derived by \citet{ramos17}. Hence, the smaller the eccentricity, the furthest from exact commensurability (i.e. $P_2/P_1=2$) a system will be. 

Resonant migration causes eccentricities to grow, a process which is balanced by eccentricity damping from the disc, leading to an equilibrium eccentricity. The equilibrium eccentricity depends on the eccentricity and semi-major axis damping times, and therefore on the disc parameters. The general formula for the equilibrium eccentricities of two planets undergoing convergent migration with different eccentricity and semi-major axis damping timescales can be found in \citet{tp19}. In the case of the 2:1 resonance, they analytically derived the equilibrium eccentricity of the inner planet as
\begin{align}
\label{eq:e1eq}
e_{\rm 1, eq}^2 = \frac{\tau_{\rm e,1}/\tau_{\rm a,2}-\tau_{\rm e,1}/\tau_{\rm a,1}}
{4\left(1+\frac{a_2 m_1}{2a_1 m_2} \right)\left(1+0.13\frac{a_2 m_1}{4a_1 m_2}\frac{t_{\rm e,1}}{t_{\rm e,2}}\right)}
\end{align}
where $i=1,2$ labels the planets with increasing distance from the star, and $\tau_{\rm e,i}$, $\tau_{\rm a,i}$, $a_i$ and $m_i$ are the eccentricity damping time, semi-major axis damping time, semi-major axis and mass of planet i, respectively. Hence, for a given pair of planets and a given set of damping times, one can use Equations (\ref{eq:resdist}) and (\ref{eq:e1eq}) to predict the departure from exact commensurability.

\subsection{Constraints and puzzles}
\label{sec:puzzles}
For the sake of discussion, we assume in this section the eccentricities of $\sim 0.07$ quoted by \citet{petigura18} for K2-24b and c. We will see later on that the eccentricities are likely to be higher than that, which would only make our argument stronger.
Assuming that the masses that we observe now for the two planets are the same masses they had when entering the resonance, Equation (\ref{eq:resdist})  suggests that in order to achieve $P_2/P_1=2.028$ (the observed period ratio of K2-24), the equilibrium eccentricity reached by $e_1$ during migration is 0.0024, a factor 30 lower than the eccentricity currently observed. At least two scenarios can resolve this discrepancy:
\begin{itemize}
\item \textit{Scenario 1}: The eccentricity and migration damping times are such that $e_1$ managed to grow to 0.06 during disc migration. According to Eq.~(\ref{eq:resdist}), the system should therefore have reached a period ratio of $P_2/P_1\simeq 2.001$ during the migration, a ratio significantly lower than what is observed. A subsequent mechanism is then required to push the planets further apart until they reach their current period ratio.
\item \textit{Scenario 2}: The eccentricity and migration damping times are such that the system naturally settles in a configuration where $P_2/P_1=2.028$ during migration. The eccentricities are smaller than what is observed today, and a subsequent mechanism is required to increase them to their current values.
\end{itemize}
In the first scenario, a mechanism is required to push the planets away from the exact commensurability. One such mechanism is tidal interactions with the star \citep[see, e.g,][]{lw12}. Since tidal forces quickly become negligible as the distance to the star increases, it is thought that this mechanism only applies to planets with period of 10 to 20 days, putting K2-24b at the limit of validity of this mechanism. More importantly, tidal interactions would damp the eccentricities, and therefore cannot be reconciled with the non-zero eccentricities of K2-24. An other mechanism involves interactions of the pair of planets with a disc of planetesimals \citep{chatterjee15}. Here too, the mechanism  leads to damping of eccentricity of the planets, and therefore does not apply to K2-24.

In the second scenario, the current period ratio is achieved during disc migrations, but eccentricities still need to be excited. This is the scenario we focus on in the remaining of the paper. The convergent migration of the pair of planets to their current period ratio is discussed in Section~\ref{sec:disc}, while the excitation of the planetary eccentricities is the purpose of Section~\ref{sec:disc3p}.

\section{Formation of the resonant pair during disc migration}
\label{sec:disc}

\subsection{Planet migration}
\label{sec:mig}
We simulate planet migration in a disc using $N$-body simulations. The equations of motion of each planet are modified to account for radial migration and eccentricity damping \citep[see, e.g.,][]{PapaloizouLarwood00}. Namely, the following acceleration:
\begin{align}
\label{eq:f_disc}
\Gamma_i= -\frac{1}{2\tau_{a,i}}\dd{\mathbf{r_i}}{t} -\frac{2}{\tau_{e,i} |\mathbf{r_i}|^2} \left(\dd{\mathbf{r_i}}{t}\cdot \mathbf{r_i} \right) \mathbf{r_i},
\end{align}
where $\mathbf{r_i}$ ($i=1,2$) is the position vector of the $i$-th planet, damps semi-major axis on a timescale $\tau_a$ and eccentricity on a timescale $\tau_e$ \citep[note that the second term on the right-hand side also gives rise to a small semi-major axis damping of order $e^2/\tau_e$; see, e.g.,][]{tt14}.

\subsection{Disc model}
In order to simulate the migration of the two planets in the disc, we need to adopt some values for the migration and damping timescales $\tau_a$ and $\tau_e$ introduced in Section~\ref{sec:mig}. In general, these timescales depend on the disc and planet properties. In particular, given the masses of K2-24b and c, they are unlikely to be in the gap-opening (i.e., Type II) migration regime. However, some of the timescale formulas derived for Earth-mass planets in the Type I migration regime may also not apply.

In this work we follow closely the model of \citet{kts18} and \citet{ks20}, since it applies well to the range of planetary masses we consider. 

We consider a disc whose surface density $\Sigma$ is given by a power-law:
\begin{align}
\Sigma=\Sigma_0 \left(\frac{r}{\text{1 au}}\right)^{-s}.
\end{align}
We adopt a small flaring for the disc scale-height, such as the disc aspect ratio $h$ is given by:
\begin{align}
h=h_0 \left(\frac{r}{\text{1 au}}\right)^{f}.
\end{align}
The disc viscosity is represented by the classical $\alpha$ parametrization, and we assume that $\alpha$ is constant throughout the disc.

The torque exerted by the disc on the planet can be scaled by the following quantity:
\begin{equation}
\Gamma_0=\left(\frac{m_p}{M_*}\right)^2 h^{-2} \Sigma r^4 \Omega_{\rm K}^2,
\end{equation}
where $\Omega_{\rm K}$ is the Keplerian frequency.
Adopting a locally isothermal disc model, the Lindblad and corotation torques normalized by $\Gamma_0$ and denoted $\gamma_{\rm L}$ and $\gamma_{\rm C}$, respectively, are:
\begin{align}
\gamma_{\rm L} &= -(2.5-0.1s+1.7\beta)b^{0.71},\\
\gamma_{\rm C} &=1.1(1.5-s)b +2.2\beta b^{0.71} - 1.4\beta b^{1.26}.
\end{align}
Here $\beta=-2f+1$ and $b=0.4h_{\rm p}/\epsilon$, where $h_{\rm p}$ is the disc aspect-ratio at the location of the planet $r_{\rm p}$, and $\epsilon$ a softening length for the planetary gravitational potential, which is taken to be 0.6 times the disc scale height at the location of the planet.

Finally, the simulations of \citet{kts18} indicate that the torque exerted by the disc on the planet is proportional to the surface density at the bottom of the gap created by the planet, rather than the unperturbed surface density. Several studies \citep[see, e.g.][]{duffell13,fung14,kanagawa15} have shown that the depth of the gap is controlled by the following parameter:
\begin{equation}
K=\left(\frac{m_{\rm p}}{M_*}\right)^2 h_{\rm p}^{-5} \alpha^{-1}.
\end{equation}
With the parameters that we consider in this paper, we have $K\sim100$ and a marginal gap can be opened, putting us in an intermediate regime between the pure Type I and Type II migration regimes.

Equipped with all these definitions, we can now give the expression for the semi-major axis damping time derived by \citet{kts18} as:
\begin{align}
\tau_a = \frac{1+0.04K}{\gamma_{\rm L}+\gamma_{\rm C}\exp{(-K/K_{\rm t})}}\tau_0(r_{\rm p}),
\end{align}
where $K_{\rm t}=20$ represents the gap depth for which the corotation torque becomes ineffective. The coefficient $\tau_0$ is defined by:
\begin{equation}
\tau_0=\frac{r^2\Omega_{\rm K} { m_{\rm p}}}{2\Gamma_0}.
\end{equation}
The eccentricity damping time is taken to be proportional to $\tau_a$:
\begin{align}
\tau_e = C_{\rm e} h^2 \tau_a,
\end{align}
where $C_{\rm e}$ is a coefficient representing uncertainties in the eccentricity damping mechanism. In the linear theory of small-mass planets, \citet{tw04} found $C_{\rm e}=1.28$. However, hydrodynamical simulations have found stronger eccentricity damping than what is suggested by the linear theory \citep[see, e.g.,][who suggested dividing $C_{\rm e}$ by 10 to better match their hydrodynamical simulations]{cresswell06}.

In order to halt migration at the observed location of K2-24b and c, we assume that once planet b reaches $a_{\rm b}=0.16~\text{au}$, all damping times start increasing with time at a rate $\exp{(t/t_{\rm slow})}$. Once $a_{\rm b}$ reaches 0.1518, the simulation stops. Several mechanisms have been studied to halt the migration at the inner edge of proto-planetary discs, which rely on the intricate physics of wave reflection and torque saturation near the inner edge \citep{tanaka02,masset06,tsang11,miranda18}. Our simple prescription gives an ideal representation of these complicated mechanisms.

\subsection{Simulation outcomes}
\label{sec:2pmig}

In this Section we present the result of 300 $N$-body simulations of planets, with the accelerations given in Eq.~(\ref{eq:f_disc}) added as extra forces. Simulations were carried on using the {\textsc \tt REBOUND} code \citep{rl12}. In Table~\ref{table:IC} we give the list of parameters that we vary, and the range of values that we chose to explore for these parameters. 

Convergent migration of the two planets is a necessary condition for them to be captured in resonance. This requires $\tau_{\rm a,2}<\tau_{\rm a,1}$. In practice, we found that this condition is satisfied for discs with surface density index $s=0.5$ or $s=1$, $h=0.025~\text{to}~0.035$,  and $\alpha=0.001~\text{to}~0.005$. We assume that the disc has a flaring index $f=1/4$. The disc surface density coefficient $\Sigma_0$ is computed using the disc mass and assuming that the disc extends from 0.25 to 50~au. The timescale for slowing down migration at the inner edge is $t_{\rm slow}=10^4~\text{yr}$. Both planets are taken to be initially on circular and coplanar orbits. Once we have drawn the initial semi-major axis of the inner planet, we take the semi-major axis of the second planet to be 1.8 times larger. The planetary masses are fixed to the ones of K2-24b and c.

\begin{table}
\caption{Initial conditions for the simulations in Section~\ref{sec:2pmig}.}
\label{table:IC}
\centering             
\begin{tabular}{l c c}
\hline 
Parameter & Notation & List of values \\ 
\hline
   Aspect ratio & $h$ & [0.025~--~0.035] \\
   Viscosity & $\alpha$ & [0.001~--~0.005] \\
   Surface density index & $s$ & 1/2 or 1 \\
   Eccentricity damping coefficient & $C_{\rm e}$  & Log-normal(-1,2) \\
   Disc mass & $M_{\rm disc}$ & [5$\Mj$--25$\Mj$] \\
   Initial position of innermost planet & $a_b$  & [0.5~au~--~1~au] \\
\hline
\end{tabular}
\end{table}

\begin{figure}
    \begin{center}
    \includegraphics[scale=0.4]{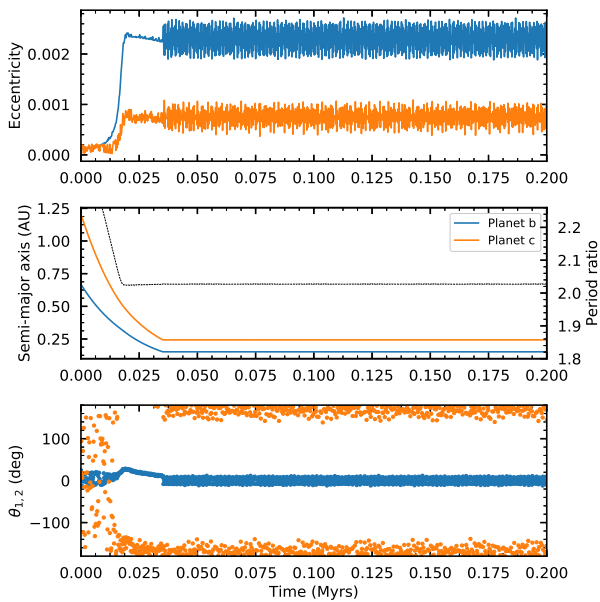}
    \caption{Disc migration and capture in the 2:1 resonance of two planets similar to K2-24b and c. From top to bottom: eccentricity, semi -major axis, and resonant angles $\theta_1$ (blue) and $\theta_2$ (orange), as defined by Eq.~(\ref{eq:res_angles}). In the middle panel, the black curve is the period ratio, whose axis is labelled on the right-hand side of the plot.}
    \label{fig:fidu_mig}
    \end{center}
\end{figure}

On Fig.~\ref{fig:fidu_mig} we show an example of two planets ending up with final periods of 20.87 and 42.31 days (period ratio 2.027), similar to the observed system. The disc has $h=0.026$, $\alpha=0.003$, $s=1$, $C_{\rm e}=0.36$ and $M_{\rm disc}=24\Mj$. This gives $\tau_{\rm a,1}=23159~\text{yr}$, $\tau_{\rm a,2}=22236~\text{yr}$, $\tau_{\rm e,1}=5.8~\text{yr}$ and $\tau_{\rm e,2}=7.4~\text{yr}$. The final eccentricities are very small, $e_1\sim 0.0024$ and $e_2\sim 0.009$,  the same as predicted by our analysis in Section~\ref{sec:puzzles}.
 
\begin{figure}
    \begin{center}
    \includegraphics[scale=0.4]{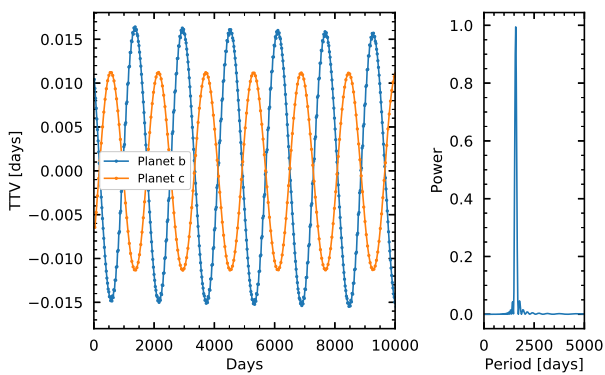}
    \caption{\textit{Left:} TTV signal for the system obtained at the end of Fig. \ref{fig:fidu_mig}. \textit{Right:} Power-spectrum of the signal, showing the pick at $\sim 1580$ days.}
    \label{fig:fidu_ttv}
    \end{center}
\end{figure}

On Fig.~\ref{fig:fidu_ttv}, we compute the TTV of the system obtained at the end of the simulation presented in Fig.~\ref{fig:fidu_mig}. The TTV signal has a period of $\sim 1582~\text{days}$, similar to the observed TTV period of K2-24. This period corresponds to the super-period (Eq.~\ref{eq:ttv_period}) for the orbital periods found in the simulation. However, the amplitude of the TTV signal is about 20 times smaller than the observed one. The small TTV amplitude is not surprising since the free eccentricity of the planets is severely damped during disc migration.

\begin{figure}
    \begin{center}
    \includegraphics[scale=0.5]{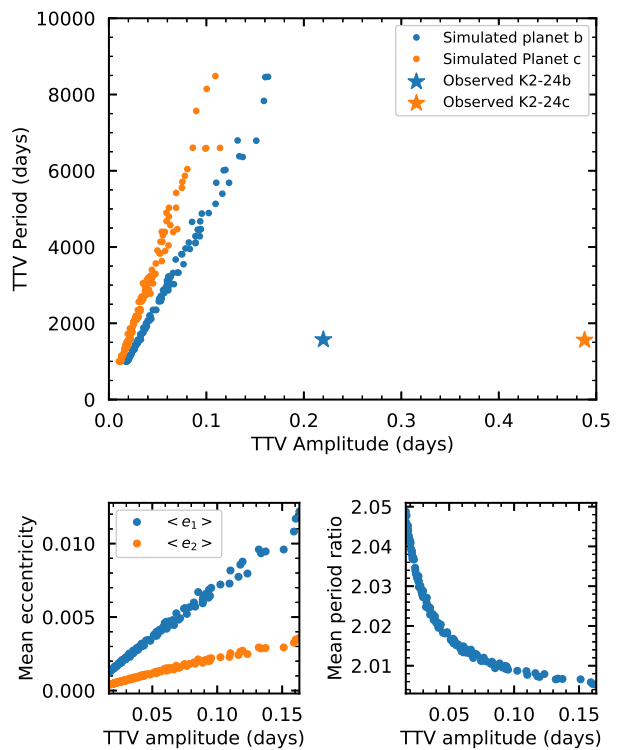}
    \caption{Outcome of all the simulations presented in Section~\ref{sec:2pmig}. \textit{Top panel:} TTV amplitude versus TTV period for all the simulated systems. The stars show the observed TTV amplitudes and periods for K2-24 \citep{petigura18}. \textit{Bottom panels:} Mean eccentricities (left) and period ratios (right) versus TTV amplitudes of the inner planet for the simulated systems.}
    \label{fig:ttv_all}
    \end{center}
\end{figure}

On Fig.~\ref{fig:ttv_all} we show the results of a series of simulations with random parameters (see Table \ref{table:IC}), for which we computed the TTV amplitude and period after convergent disc migration and capture in the 2:1 resonance. Although some systems achieve high TTV amplitude, they do so at the expense of having a very long TTV period. The fact that the TTV period increases with amplitude can be understood by combining Eqs.~(\ref{eq:ttv_period}) and (\ref{eq:resdist}): the TTV period will increase linearly with eccentricity, and therefore with TTV amplitude \citep{lithwick12,deck16}. This is confirmed by the bottom panels of Fig.~\ref{fig:ttv_all}, where we also show the averaged eccentricity of each planet, and their averaged period ratio, versus the TTV amplitude of planet b (the averages are computed over the time in which the planets have stopped migrating and reach an equilibrium eccentricity). TTV amplitude (and therefore TTV period) increases with eccentricity, but decreases with period ratio.

We note that all the resonant pairs that we form through disc migration end up with the inner planet having the largest TTV, as seen in Fig.~\ref{fig:fidu_mig}. This is the opposite of what is observed in K2-24 (star symbols in Fig.~\ref{fig:ttv_all}), where the outer planet has the largest TTV amplitude. This may indicate that whatever process acted to excite the planetary eccentricities was more efficient at exciting the outer planet.

\section{Disc-induced precession and eccentricity excitation}
\label{sec:disc3p}

It is clear from the previous section that, although it is possible to generate K2-24's current periods during disc migration, a subsequent mechanism is necessary to excite its eccentricities to a level that can explain the observed TTVs. In this section we present one such possible scenario.

\subsection{Disc-induced precession}

In this section we consider that an inner cavity has been cleared in the disc. In this cavity orbit K2-24b and c, and also K2-24d, the third planet whose possible detection is mentioned by \citet[][see Section \ref{sec:review}]{petigura18}. At this point, since all the planets are in the cavity, we assume that they no longer migrate. However, the disc still exerts a gravitational torque on the planets.

Following \citet{tl19}, the disc causes planets interior to it to precess at the rate
\begin{equation}
\label{eq:disc_prec}
\dot{\varpi}=\frac{3}{4}\frac{M_{\rm loc}}{M_*}\left(\frac{a_{\rm p}}{r_{\rm in}}\right)^3\Omega_{\rm p}~\hat{\omega},
\end{equation}
where $a_{\rm p}$ and $\Omega_{\rm p}$ are the semi-major axis and Keplerian frequency of the planet. In addition, $r_{\rm in}$ is the inner radius of the disc, and  $M_{\rm loc}=2\pi\Sigma r_{\rm in}^2$ is the local mass of the disc at the inner radius. Finally, $\hat{\omega}$ is a dimensionless integral whose expression can be found in \citet{tl19}.

We assume that the cavity in the disc is such that $r_{\rm in} \gtrsim a_{\rm d}$, the semi-major axis of planet~d. Hence in our case, the disc-induced precession is mostly going to affect planet~d. Assuming that the disc slowly dissipates over time, e.g. due to photoevaporation, the precession rate of planet~d is going to slowly change over time. As shown by \citet{lw11} an outer massive planet with a varying precession rate can cause the planets interior to it to cross a secular resonance, which would excite their eccentricities. 

In order to simulate the effect of the disc dispersal, we assume that $\dot{\varpi}$ starts from an initial value $\dot{\varpi}_0$ and decays with time as
\begin{equation}
\dot{\varpi}(t)=\dot{\varpi}_0 \me^{-t/t_{\rm disp}},
\end{equation}
where $t_{\rm {disp}}$ is the dispersal timescale. This simple law is meant to represent whatever process is dispersing the gaseous disc in the later phase of its life, e.g., photo-evaporation or magnetic winds.

The disc-induced precession described by Eq.~(\ref{eq:disc_prec}) can be readily implemented in {\textsc \tt REBOUNDx} \citep{tamayo19}.

\subsection{Results}
We ran 600 $N$-body simulations that include 3 planets, the precession term given by Eq.~(\ref{eq:disc_prec}) and relativistic corrections. The initial conditions for K2-24b and c are the final parameters of Fig.~\ref{fig:fidu_mig}. The mass and semi-major axis of K2-24d are drawn uniformly from the range of possible values quoted by \citet{petigura18}, i.e. [$40~M_{\oplus}$--$68~M_{\oplus}$] and [1.1~au--1.21~au], respectively. Its eccentricity is drawn from a uniform distribution between 0 and 0.3. Its mean longitude and argument of pericentre are drawn from a uniform distribution between 0 and $2\pi$. The three planets and the disc are assumed to be coplanar. The inner radius of the disc is drawn from a uniform distribution between 1.2 and 2~au, while its mass varies uniformly from 5 to $25~\Mj$. The disc surface density is a power-law with index $s=1$. The dispersal timescale of the disc varies uniformly between $10^{5}$ and $3\times 10^{5}~\text{yrs}$.

With this particular set of parameters, we found that eccentricity excitation of the inner pair was a common outcome of our simulations, with 50\% of the simulations resulting in $e_{\rm b}>0.05$, and 5\% resulting in $e_{\rm b}>0.1$.

\begin{figure}
    \begin{center}
    \includegraphics[scale=0.5]{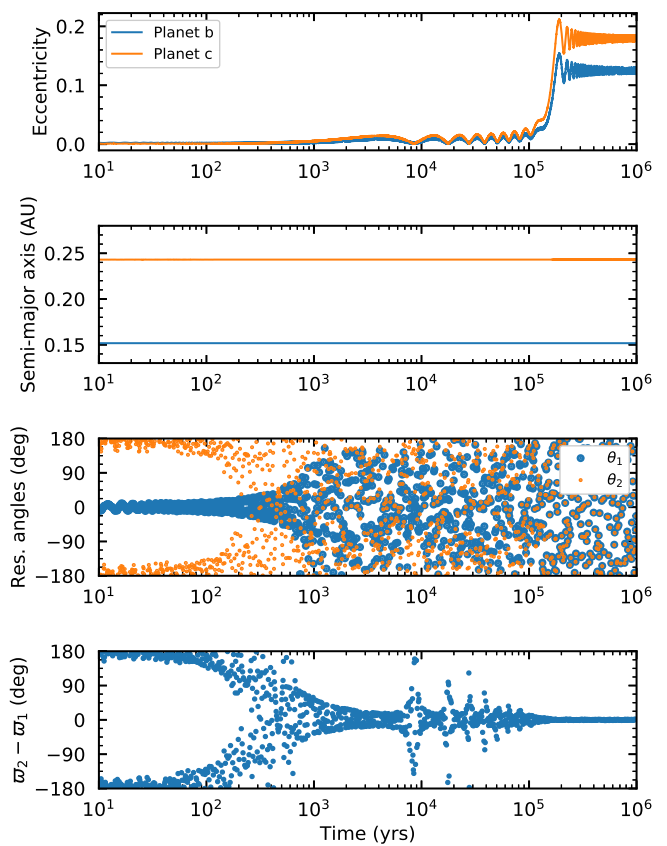}
    \caption{Time evolution of the inner pair K2-24b and c, when perturbed by an outer precessing planet and an evaporating disc. From top to bottom: eccentricity, semi-major axis, resonant angles (see Eq.~\ref{eq:res_angles}) and difference of argument of pericentres. }
    \label{fig:3p}
    \end{center}
\end{figure}

On Fig.~\ref{fig:3p} we show an example of evolution of K2-24b and c, as an evaporating disc causes K2-24d to precess at a time-varying rate. The outer planet has the following properties: $m_3=61.2~M_{\oplus}$ $a_3=1.12~\text{au}$ and $e_3=0.25$. The disc extends from $\rin=1.8~\text{au}$ to $\rout=30~\text{au}$, and initially contains a mass of $20~\Mj$. This mass decays exponentially on a timescale $t_{\rm { disp}}=2\times10^{5}~\text{years}$. 
As the disc mass decays in time, it alters the precession of the outer planet, which in turn secularly excites the eccentricity of the inner pair,  while maintaining their semi-major axis (and therefore period-ratio) constant. The final eccentricities of K2-24b and c are 0.12 and 0.17. Their  resonant angles do not librate any more, and the pair is locked in a mutual apsidal precession around 0. The orbit of K2-24d, not shown here, remains largely unaffected. In that sense, K2-24d merely acts as a messenger to propagate a time-varying secular forcing from the disc onto the interior planets.

\begin{figure}
    \begin{center}
    \includegraphics[scale=0.5]{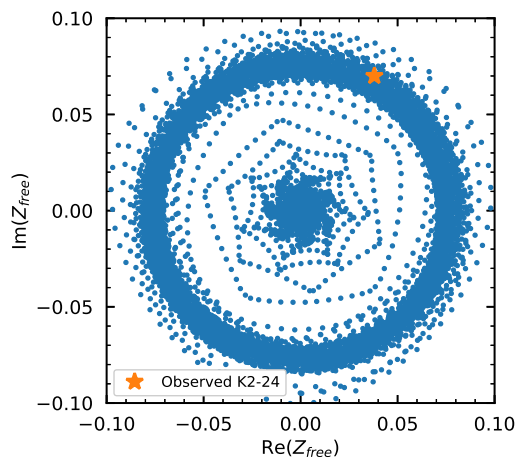}
    \caption{Trajectory of $Z_{\rm free}$ in the real--imaginary plane (see Eq.~\ref{eq:zfree}) for the system shown in Fig.~\ref{fig:3p}. The orange star marks the current observed location of K2-24 in that plane (see Eq.~\ref{eq:zfreek224}),}
    \label{fig:3p_Z}
    \end{center}
\end{figure}

On Fig.~\ref{fig:3p_Z} we show the evolution of $Z_{\rm free}$ (see Eq.~(\ref{eq:zfree})) for the simulation presented in Fig.~\ref{fig:3p},  in the $\text{Re}(Z_{\rm free})$--$\text{Im}(Z_{\rm free})$ plane. At the beginning of the simulation, the free eccentricity of the pair is small since it has been damped during disc migration, and it is therefore concentrated around $(0,0)$. As the eccentricity is excited, the trajectory expends, until it settles and encompasses the current value of K2-24, indicated by an orange star in this plot.

\begin{figure}
    \begin{center}
    \includegraphics[scale=0.4]{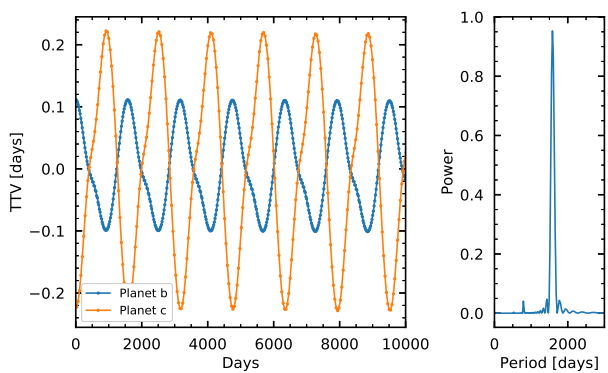}
    \caption{TTV computed at the end of the simulation shown in Fig.~\ref{fig:3p}. The amplitude of each TTV signal, as well as their periods, are the same as those observed in K2-24.}
    \label{fig:3p_tv}
    \end{center}
\end{figure}

\begin{figure}
    \begin{center}
    \includegraphics[scale=0.32]{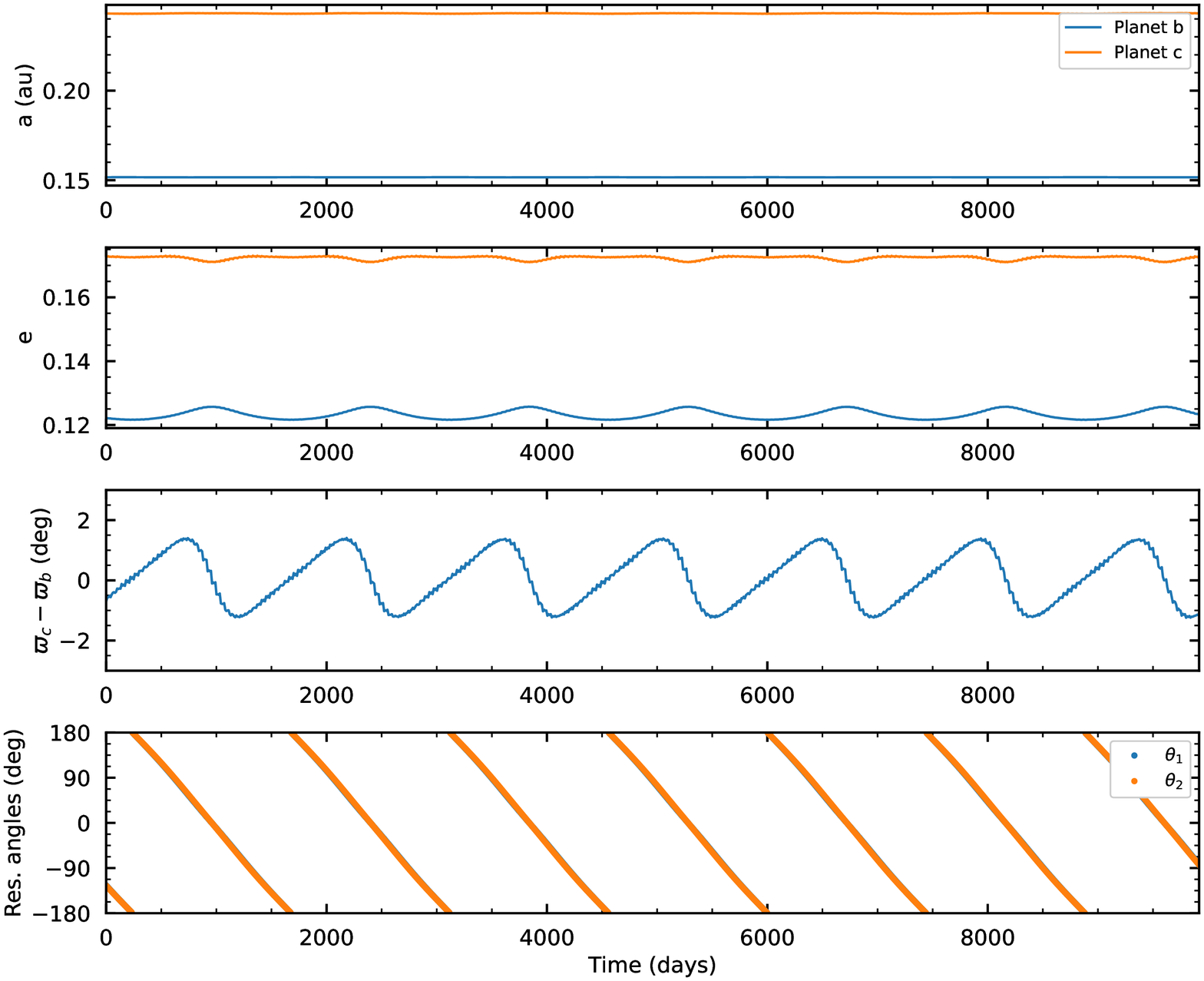}
    \caption{Orbital evolution of planets b and c during the TTV simulated on Fig.~\ref{fig:3p_tv}. From top to bottom: semi-major axis, eccentricities, difference of pericenter arguments $\varpi_{\rm c}-\varpi_{\rm b}$, and resonant angles $\theta_1$ and $\theta_2$. }
    \label{fig:3p_orb}
    \end{center}
\end{figure}

We can now compute the TTVs of the system at the end of the simulation shown in Fig.~\ref{fig:3p}. We show the TTVs on Fig.~\ref{fig:3p_tv} (with corresponding orbital evolution on Fig.~\ref{fig:3p_orb}). As expected, the signal maintains the same period as before, since the orbital periods have not changed. However the amplitude of the signal is now much stronger than it was at the end of the disc migration phase (see Fig.~\ref{fig:fidu_ttv}). In this particular example, the amplitudes are 0.26 and 0.48 days for planets b and c, respectively, in good agreement with the observed values reported by \citet{petigura18}. At the end of the disc-induced migration phase presented in Section~\ref{sec:disc}, the TTV amplitude of the inner planet was smaller than that of the outer planet, which was at odds with observations. This is no longer the case. Finally, we note that the periodogram associated with the TTV signal of Fig.~\ref{fig:3p_tv} shows a small secondary peak at a period of 790 days, half of the main period. As noted by \citet{hadden16}, for a pair of planets near the 2:1 MMR, there exists a second-harmonic signal associated with the 4:2 MMR, whose frequency is twice that of the main signal, and whose amplitude is smaller by a factor $\sim e$. This is the secondary peak that we see on the periodogram of Fig.~\ref{fig:3p_tv}.

On Figure~\ref{fig:3p_orb} we show the orbital evolution of planets b and c during the TTV simulated on Fig.~\ref{fig:3p_tv}. We do not show the orbital evolution of planet~d since it is barely affected by the inner pair and its orbital elements do not significantly vary over the time of integration. We see that the orbital elements vary on the same period as the TTVs, and that the system is no longer in resonance. Since $\varpi_{\rm b}\simeq \varpi_{\rm c}$, the two resonance angles $\theta_1$ and $\theta_2$ almost perfectly overlap.

\section{Discussion}
\label{sec:discussion}

\subsection{Main assumptions}

In order to reproduce the observed properties of K2-24, we built a scenario in two distinct parts: In Section~\ref{sec:disc} we focused on the migration of K2-24b and c in a disc, in order to reproduce their observed period ratio. We did not take into account the presence of the third planet at this stage. We then took the final outcome of one of our simulations, which had the correct period ratio, and used it as the starting point for an other set of simulations. This set of simulations, which we presented in Section~\ref{sec:disc3p}, assumed that the disc had partially depleted and formed an inner cavity, and that in this cavity also orbited a third planet. We therefore need to justify two assumptions: i) why did we not take into account planet~d in Section~\ref{sec:disc}, and ii) why did we start the simulations of Section~\ref{sec:disc3p} with a cavity already carved?

Regarding the first assumption, it is likely that K2-24d formed further out than its current position and migrated in the disc just like K2-24b and c did, but did not have time to reach the inner regions of the disc before it started evaporating. As long as K2-24d did not migrate faster than the inner pair, so to prevent resonant capture, we assume that the disc would suppress any planet-planet interaction between K2-24d and the inner pair. Therefore their dynamical evolutions would be largely disconnected and the influence of K2-24d can be ignored.

We now turn to discuss the second assumption. At the end of Section~\ref{sec:disc}, we were left with two planets which had stopped migrating at about 0.15 and $0.24~\text{au}$. This implies that the gaseous disc extended to roughly the same location. However we started the simulations of Section~\ref{sec:disc3p} with a disc inner radius at $1.8~\text{au}$, implying that the disc had cleared an inner cavity. We did not include the intermediate phase where the disc's inner edge expends from the vicinity of K2-24c to beyond the orbit of K2-24d. As we point out in the next section, it is likely that some instabilities altered the orbit of K2-24d very early on, making the full evolution rather tedious to simulate. Hence we take the simplifying assumption to separate the evolution history in two distinct parts.

\subsection{Orbital properties of K2-24d}
Our mechanism for exciting the eccentricities of planets b and c up to their observed values relies on a secular resonance crossing provoked by the time-varying precession of planet~d, induced by an outer disc. As shown by \citet{lw11}, the strength of this mechanism increases with increasing eccentricity of planet~d. In general, we found that eccentricities between 0.2 and 0.3 were adequate. This raises the question of how planet~d achieved this eccentricity so early on. Several works have shown that disc-planet interactions can increase the eccentricity of giant, gap-opening planets \citep[see, e.g.][]{papaloizou01,gs03,dlb06,rice08,to16,ragusa18,muley19}. However, it is likely that K2-24d is not massive enough for these mechanisms to apply, and it is also not clear whether planet-disc interactions can excite planet eccentricities to such high values. More promisingly, early instabilities of multi-planet systems during disc-planet interactions have been shown to lead to planet-planet scattering \citep{marzari10,moeckel12,lega13,rosotti17}. This scattering can lead to the ejection of one or more planets, leaving the remaining planet(s) eccentric. If such event occurred in the outer planetary system of K2-24, it provides a justification for choosing a non-zero eccentricity of K2-24d.

We note that in principle, the two mechanisms discussed above (eccentricity growth due to planet-disc interactions or planet-planet scattering) could also excite the free eccentricity of the inner pair, without having to rely on the existence of a third planet at all. However there is no evidence that planet-disc interaction can increase the eccentricity of planets in the mass range of K2-24b and c. On the other hand scattering events result from a kick in orbital energy, and therefore would not preserve the semi-major of the inner pair.

\subsection{Disc properties}
In Section~\ref{sec:disc}, we have assumed that the two planets undergo smooth migration in a non-turbulent disc. However, \citet{petigura18} argued that the eccentricity of K2-24b and c could be the result of stochastic migration in a turbulent disc \citep{laughlin04,ogihara07,adams08}. We implemented the stochastic forces suggested by \citet{laughlin04,ogihara07} into our $N$-body code to test whether stochastic migration could reproduce both the eccentricities and period ratio of K2-24b and c. We found that the turbulent forces had little effect on the outcome once the two planets are captured in a MMR, and therefore we discarded that scenario and did not include turbulent forces in subsequent simulations.

When simulating the migration of the two planets in the disc, we have implemented damping formulas from \citet{kts18,ks20}. Although these formulas were used by \citet{ks20} for the migration of two planets, it is worth pointing out that they were derived from hydrodynamical simulations consisting of one planet in a disc. The presence of a second planet will change the disc structure, and the creation of a common gap could slow down the migration rate. In addition, the disc torque may be modified by overlapping spiral density waves, as noted by \citet{broz18}. The effect of the disc's self-gravity can also modify the migration rate and the occurence of resonant capture \citep{ataiee20}.

\subsection{Long-term stability}
Finally, it remained to verify that the system obtained in Fig.~\ref{fig:3p}, with its three eccentric planets, is stable over long timescales. We carried a $N$-body simulation over 100 Myr and observed no destabilisation of the system. Since this does not guarantee the stability of the system over the age of the system \citep[the star is several Gyr old; see][]{petigura16}, we also computed Mean Exponential Growth factor of Nearby Orbits stability maps \citep[MEGNO, which are readily implemented in {\textsc \tt REBOUND}, see also][]{cincotta00} where we varied the semi-major axis, eccentricity and mass of K2-24d on a grid of values. The grid extended from $a_{\rm d}=1.1$ to $a_{\rm d}=1.21 \,\text{au}$, $e_{\rm d}=0.21$ to $e_{\rm d}=0.27$, and $m_{\rm d}=40M_{\oplus}$ to $m_{\rm d}=68M_{\oplus}$. The resolution of the grid was $20\times20\times5$. These simulations all yielded MEGNO values close to~$2$ and therefore did not indicate any evidence for chaos in the system.

\section{Conclusion}
\label{sec:conclusion}

In this paper we have studied the impact of disc-induced migration on the subsequent TTV signal produced by two planets near the 2:1 MMR. We have focused on the K2-24 system, and shown that disc-induced migration can reproduce the correct ratio of orbital period of the innermost two planets. However, systems formed with the correct period ratio exhibit a TTV signal that is 20 times smaller than the observed TTVs. Hence, in the particular case of K2-24, we suggest that disc-driven migration alone cannot account for the observed properties of the system. We have proposed that an additional mechanism came into play in the late phase of planet-disc interactions. We have found one such possible mechanisms, involving the crossing of a secular resonance induced by a third outer planet, whose precession rate is varying due to the evaporating gaseous disc. In this scenario, the planets remain in their observed period ratios, and reach eccentricities that are compatible with observations. The period and amplitude of our simulated TTVs match the observed one. 

More generally, we have conducted the first study of the impact of disc migration on the amplitude and period of TTV signals. Fig.~\ref{fig:ttv_all} is promising because it shows that if resonant pairs were formed during disc migration and remained unaltered after that, their TTV signal should lie on a linear track in the amplitude--period diagram. Although this track will be different for each system, direct $N-$body simulations with migration forces can show whether or not an observed system lies on this track, and shed new light on its formation history.

\begin{acknowledgements} 
The work of JT is supported by a Fonds de la Recherche Scientifique – FNRS  Postdoctoral Research Fellowship.
Computational resources have been provided by the PTCI (Consortium des Équipements de Calcul Intensif CECI), funded by the FNRS-FRFC, the Walloon Region, and the University of Namur (Conventions No. 2.5020.11, GEQ U.G006.15, 1610468 and RW/GEQ2016).
The research done in this project made use of the SciPy stack \citep{scipy}, including NumPy \citep{numpy} and Matplotlib \citep{matplotlib}, as well as Astropy,\footnote{\url{http://www.astropy.org}} a community-developed core Python package for Astronomy \citep{astropy:2013, astropy:2018}. Simulations in this paper made use of the REBOUND code which is freely available at \url{http://github.com/hannorein/rebound}.
\end{acknowledgements}

\bibliographystyle{aa}
\bibliography{biblio}

\label{lastpage}
\end{document}